\documentclass[aps,twocolumn,superscriptaddress,longbibliography]{revtex4-2}
\usepackage{graphicx,amsmath,amsfonts,amssymb,bm,algorithm,algpseudocode,booktabs}
\usepackage[colorlinks=true, urlcolor=blue, linkcolor=blue, citecolor=blue]{hyperref}

\begin{document}

\title{From Paper to Program: Specification-Guided Implementation of Quantum Many-Body Algorithms with Large Language Models}

\author{Yi Zhou}
\email{yizhou@iphy.ac.cn}
\affiliation{Institute of Physics, Chinese Academy of Sciences, Beijing 100190, China}

\date{\today}

\begin{abstract}
Turning scientific papers into reliable code requires making implicit computational knowledge explicit. We introduce and evaluate a human-supervised workflow in which large language models develop and review technical specifications before implementing quantum many-body algorithms. The specifications encode conventions, procedures, scaling requirements, and validation gates for mathematical consistency, physical accuracy, and production-scale performance. For the density-matrix renormalization group, all 16 tested specification--implementation model pairings succeed, compared with 6 of 13 direct attempts; two further attempts succeed with a prose specification. For Pfaffian conversion of Hartree--Fock--Bogoliubov states to matrix product states, on eleven matched model--harness configurations, four self-paired workflow attempts pass and no direct attempts do; across all phases, 11 of 26 workflow attempts pass the audit and validation criteria. Cross-model transfer is asymmetric: one implementation model succeeds with all four specifications written by other models, whereas all four reverse-direction attempts fail. A consolidated specification informed by these experiments supports successful implementations in three previously unsuccessful model--harness configurations. Technical specifications therefore guide scalable implementation and help distinguish paper-to-code ambiguity from limitations in implementation capability. We synthesize these findings into the \emph{Paper-to-Program Many-Body} protocol, which turns implementation experience into reusable review criteria, validation gates, and reporting practices. Publicly archived prompts, transcripts, specifications, code, and validation records support independent verification and extension to other many-body algorithms.
\end{abstract}

\maketitle

\section{Introduction}
\label{sec:intro}

Scientific papers communicate methods compactly to expert readers, relying on shared conventions and leaving computational details to expert judgment. Producing reliable code requires index order, data layout, numerical conventions, and scaling choices to be explicit and mutually consistent. Large language models (LLMs) can accelerate this process by combining scientific interpretation with code generation. The practical question is how to organize these capabilities so that the resulting code faithfully realizes the method, scales to relevant system sizes, and passes independent scientific validation.

We investigate a human-supervised, agent-driven workflow that places reviewed technical specifications between source literature and implementation. Each specification makes the method computationally explicit by recording tensor shapes, contraction sequences, gauge choices, fermionic signs, memory requirements, and validation tests. We refer to this process as \emph{knowledge externalization}. The specifications serve a constructive role by guiding implementation and a diagnostic role when reused across agents: a common computational reference lets differences in outcome be traced to the source presentation or to the implementation agent.

Recent benchmarks and agentic workflows address complementary aspects of LLM-assisted scientific programming. SciCode evaluates scientific coding problems curated by researchers~\cite{Tian2024SciCode}, while PaperBench tests the reproduction of machine-learning research~\cite{Starace2025PaperBench}. PaperCoder uses planning, analysis, and code-generation agents to develop repositories from machine-learning papers~\cite{Seo2026Paper2Code}. In computational physics, structured multistep prompting has been used to reconstruct Hartree--Fock derivations~\cite{Pan2025ManyBodyLLM}, while curated documentation and agent decomposition have supported tensor-network simulations~\cite{Li2026AutonomousQuantum}. Our study makes reviewed technical specifications themselves an object of evaluation. We test their transfer across implementation models and refine them using evidence from physical and computational validation.

Quantum many-body algorithms provide a demanding and physically informative testbed for specification-guided implementation. Tensor-network methods~\cite{Verstraete2008,Orus2014,Cirac2021}, including the density-matrix renormalization group (DMRG)~\cite{White1992,White1993,Schollwock2011}, manipulate matrix product state (MPS) representations~\cite{Fannes1992,Ostlund1995,PerezGarcia2007} through structured multi-index contractions. Their implementation requires consistent index ordering, well-defined canonical gauges, and control of memory growth. Ground-state energies, entanglement, and nonlocal order provide complementary physical benchmarks for the resulting states. These features jointly test mathematical consistency, computational scalability, and physical accuracy.

We study two algorithms with contrasting computational structures and source presentations. DMRG is reconstructed from Schollw{\"o}ck's approximately 100-page review~\cite{Schollwock2011} and serves as a calibration case: its detailed pedagogical source describes variational sweeps and local effective-Hamiltonian eigenproblems. By contrast, Pfaffian conversion of Hartree--Fock--Bogoliubov (HFB) states to MPS is drawn from a five-page Letter~\cite{Jin2022} and serves as a compact-source stress test. Constructing the MPS tensors from overlaps of Bogoliubov vacua requires consistent fermionic canonicalization, parity conventions, and Pfaffian ordering. The conversion connects paired-fermion mean-field states to tensor-network calculations and Gutzwiller-projected parton constructions~\cite{Jin2020,Jin2022}, while the resulting MPS can seed Gutzwiller-guided DMRG~\cite{Jin2021}. Validated implementations therefore feed directly into subsequent many-body calculations.

Established libraries such as ITensor~\cite{ITensor2022} and TeNPy~\cite{TeNPy2018} provide mature support for tensor-network calculations, while TeMFpy implements fermionic mean-field-to-MPS conversion~\cite{HilleSzabo2025TeMFpy}. To study the paper-to-program process directly, we use a closed-world protocol in which agents reconstruct standalone implementations from source literature or derived technical specifications. The agents use the supplied documents, their general scientific knowledge, and local numerical tools without retrieving existing implementations. Human supervision evaluates runtime behavior and physical observables, while the agents carry out implementation and debugging. This division of labor links agent-driven development to scientific assessment through the staged protocol described in Sec.~\ref{sec:methodology}.

The experiments yield three connected findings. First, all 16 tested DMRG specification--implementation model pairings succeed, compared with 6 of 13 direct attempts; two additional attempts succeed with a prose specification, showing that technical guidance remains effective without mathematical typesetting. Second, for Pfaffian-MPS, the same eleven model--harness configurations yield four passing self-paired workflow attempts and no passing direct attempts, and 11 of 26 workflow attempts pass across all phases. Cross-model transfer is asymmetric: GPT~5.5 under Codex succeeds with all four tested specifications written by other models, whereas none of those four models succeeds with GPT~5.5's specifications. Third, a consolidated specification incorporating lessons from earlier runs supports successful implementations in three previously unsuccessful model--harness configurations.

Together, these findings show that specifications serve constructive and diagnostic roles and support a two-bottleneck picture of AI-assisted scientific programming: explicit computational guidance reduces paper-to-code ambiguity, while implementation agents contribute the reasoning, debugging, and computational organization needed to realize the method. Feedback from implementation and validation then refines both the specifications and the workflow, turning experience from individual runs into reusable review criteria, validation gates, and reporting practices. The \emph{Paper-to-Program Many-Body} protocol operationalizes this cycle as a reusable agent skill. A public archive of prompts, transcripts, specifications, code, and validation results~\cite{yizhou_dmrg_llm_2026} supports independent verification and extension to other many-body algorithms.

\section{Specification-Guided Workflow and Evaluation Protocol}
\label{sec:methodology}

\begin{figure*}[tb]
    \centering
    \begingroup
    \setlength{\unitlength}{0.98\textwidth}
    \setlength{\fboxsep}{0pt}
    \begin{picture}(1,0.545455)
        \put(0,0){\includegraphics[width=\unitlength]{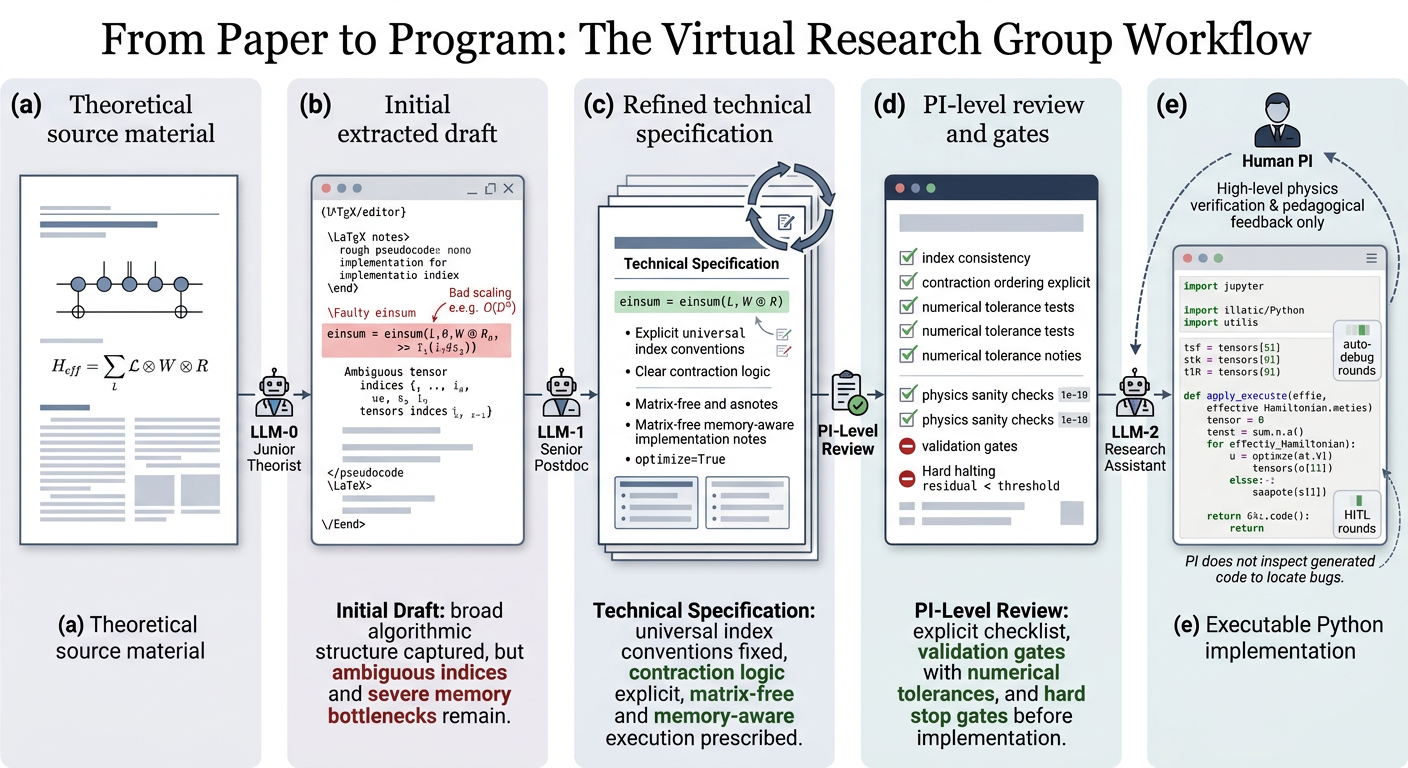}}
        \put(0.651,0.192){\colorbox{white}{\parbox[b][0.027\unitlength][c]{0.110\unitlength}{\sffamily\fontsize{5}{6}\selectfont\raggedright Hard stop gate\\$r\geq\varepsilon$}}}
    \end{picture}
    \endgroup
    \caption{\textbf{The \emph{Paper-to-Program} workflow.} \textbf{(a)}~Source literature provides the theoretical starting point. \textbf{(b)}~The extraction agent produces an initial theory document. \textbf{(c)}~The specification agent makes computational conventions, contraction sequences, and scaling requirements explicit. \textbf{(d)}~An LLM reviewer applies the PI-level review checklist and defines validation and stop gates. For a test with acceptance criterion $r<\varepsilon$, a stop gate halts dependent steps whenever $r\geq\varepsilon$. \textbf{(e)}~The implementation agent generates, tests, and debugs the code. Throughout the workflow, the human PI provides scientific supervision and feedback on runtime behavior and physical observables.}
    \label{fig:workflow}
\end{figure*}

The expanded workflow assigns complementary responsibilities to four agent roles: theory extraction (LLM-0), technical specification (LLM-1), specification review (LLM reviewer), and code implementation (LLM-2). These functional roles may be filled by the same or different foundation models. A human principal investigator (PI) selects the scientific task, supervises the process, and evaluates physical diagnostics. Figure~\ref{fig:workflow} presents this division of labor, termed a ``virtual research group'' in the archived prompts. Each agent role has a defined task and a reviewable output, making the progression from scientific interpretation to implementation explicit.

The protocol evolved through two successive workflow generations. The DMRG study used the earlier three-stage sequence of theory extraction, technical specification, and implementation. Experience from the initial Pfaffian-MPS runs motivated an expanded workflow with dedicated specification review and structured implementation reporting to address fermionic signs, gauge consistency, observable provenance, and computational scaling. The remainder of this section describes the expanded protocol. Section~\ref{sec:pfaffian-workflow} organizes the Pfaffian-MPS attempts by specification strategy, while run-local documents preserve the available review and refinement history. As detailed in Sec.~\ref{sec:pfaffian-phase3}, the consolidated specification continued to be revised in response to implementation feedback.

\subsection{Theory Extraction and Specification Development}

The extraction agent receives the source paper and produces an initial theory document, \texttt{latex-spec-LLM-0.tex}. This draft organizes the Hamiltonians, canonical-form relations, contraction rules, algorithmic steps, and benchmark definitions needed for implementation. The extraction prompt directs the agent to resolve cross-references, expand compressed equations, and record any notational clarifications it introduces.

The specification agent reviews the draft equation by equation for computational consistency and implementation readiness, producing \texttt{latex-spec-LLM-1.tex} through autonomous refinement cycles. At the end of each cycle, the agent classifies the draft as \textsc{coding-ready} or \textsc{not coding-ready} and records any issues requiring further revision. A \textsc{coding-ready} classification designates a candidate for implementation; in the expanded Pfaffian-MPS workflow, the candidate proceeds to the dedicated review described below. The resulting specification organizes implementation guidance into four elements:
\begin{itemize}
    \item \textbf{Conventions and shapes:} index ranges, tensor-leg order, basis conventions, gauge choices, and array dimensions.
    \item \textbf{Operations:} contraction sequences specified by explicit \texttt{einsum} strings or \texttt{tensordot} axes, together with algorithmic pseudocode and subroutine inputs and outputs.
    \item \textbf{Computational architecture:} memory costs, choices between stored and on-demand intermediates, matrix-free operations, and reuse of intermediate contractions.
    \item \textbf{Validation:} tests of mathematical identities, parity, normalization, eigenvalue pairing, and reconstruction, each with explicit numerical tolerances.
\end{itemize}
The human supervisor defines the scientific problem and high-level requirements, while the specification agent translates them into detailed computational content, including contraction formulas and algorithmic procedures.

\subsection{Specification Review and Validation Gates}
\label{sec:pi-review}

In the expanded Pfaffian-MPS protocol, a dedicated LLM reviewer examines each candidate specification before implementation. This archived stage, labeled \emph{PI-Level Review}, uses a PI-style prompt to assess whether the implementation requirements are complete and internally consistent. The review supplements the specification agent's coding-readiness assessment with checks motivated by ambiguities observed in larger-system tests and defines validation and stop gates for execution.

The reviewer applies an algorithm-specific checklist covering operator order, canonical forms, gauge fixing, fermionic signs, tensor shapes, and computational reuse. Each item is classified as \textsc{clear}, \textsc{needs work}, or \textsc{missing}; unresolved items return the specification for further refinement. The review establishes \emph{validation gates}, which define required checks and acceptance criteria, including numerical tolerances where applicable, and \emph{stop gates}, which prevent dependent steps from proceeding when a prerequisite fails. Both gate types become part of the specification supplied to the implementation agent: the reviewer defines the requirements, and the implementation agent executes and reports the corresponding tests.

Validation and stop gates organize implementation as a sequence of verifiable stages. For example, canonical-form reconstruction must meet its prescribed tolerance before agreement with a downstream energy benchmark can count as validation evidence. If reconstruction fails, a stop gate requires the agent to resolve the inconsistency and repeat the check before proceeding. Stop gates thus trigger repair and revalidation within a run, while the final outcome depends on the full set of required checks. This staged structure focuses debugging and ties physical benchmarks to their computational prerequisites.

\subsection{Implementation, Debugging, and Reporting}
\label{sec:llm2}
\label{sec:reporting}

Across the experiments, the implementation agent translates the supplied source document or reviewed specification into Python code and executes the required validation checks. Runs use the Codex, ClaudeCode, OpenCode, and KimiCode agentic harnesses. We distinguish two debugging loops. During an \emph{auto-debug round}, the agent executes the code, interprets a runtime error or failed check, and revises the implementation autonomously. During a \emph{human-in-the-loop} (HITL) round, the human PI supplies a traceback or physics-level diagnostic after the initial code delivery. Reported HITL counts include only these subsequent feedback rounds and exclude the initial implementation prompt.

Human feedback addresses runtime behavior and physical observables, while the agent remains responsible for computational diagnosis and code revision. Representative feedback concerns a shape-mismatch traceback, collapsed entanglement entropy, a reported MPS energy numerically identical to the exact correlation-matrix energy, or a production-size benchmark that did not complete. During HITL debugging, the PI does not inspect source code to locate bugs, identify incorrect lines, provide corrected contractions or sign formulas, or write replacement routines. The transcripts preserve these interventions, so the scope of scientific supervision can be examined alongside implementation outcomes.

In the Pfaffian-MPS workflow, the implementation agent also produces a structured report, \texttt{stage3-report.md}. It records code organization, algorithmic choices, numerical tolerances, auto-debug and HITL rounds, benchmark values, runtime and memory measurements, deviations from the specification, and remaining issues. Each test is labeled as passed, failed, or not run, and the report states how each observable was computed. This documentation supports the audit of production-scale validation by preserving both numerical results and their measurement provenance, including whether reported MPS energies were obtained by contracting the constructed state. Across runs, the reports provide evidence for outcome classification and reveal computational guidance that can be incorporated into later specifications.

\subsection{Experimental Design and Run Conditions}
\label{sec:experimental-design}

Each experimental condition is characterized by its input document, specification and implementation models, agent roles, execution harness, implementation constraints, and benchmark task. Direct baselines supply the source paper to the implementation model, whereas specification-guided runs supply an intermediate technical specification. In both cases, the prompts state the high-level physics and scaling requirements. The DMRG experiments evaluate a $4\times4$ grid of specification--implementation model pairings together with a prose-specification comparison. The Pfaffian-MPS experiments evaluate same-model pairing, cross-model specification transfer, and implementation from a consolidated specification. The case-study sections provide the detailed benchmark settings for each condition.

All experiments use a closed-world implementation setting. Agents work from the supplied source literature or technical specification and may use their general scientific knowledge and local execution tools, but not external web search or retrieved implementations. Numerical code uses NumPy and SciPy, with Matplotlib for plotting. These source and dependency boundaries focus the evaluation on reconstructing standalone programs from the supplied scientific descriptions.

Debugging budgets follow the archived protocol for each experimental condition. DMRG direct-baseline attempts used a stopping guideline of approximately 20 HITL rounds without convergence, with earlier termination when a model declared that it could not complete the task; one unsuccessful Claude run continued to 32 rounds. Pfaffian-MPS direct baselines allowed up to 30 auto-debug rounds, and each prose-specification attempt allowed 20 HITL rounds. The remaining specification-guided runs had no preset HITL limit, so their round counts reflect the errors encountered and the autonomous-debugging behavior of the harness. The comparisons therefore characterize the recorded development protocols, including their input documents, execution environments, and supervision histories.

\subsection{Validation Criteria and Outcome Accounting}
\label{sec:protocol}

A run is classified as successful only if it satisfies three requirements: executable code that respects the dependency and provenance constraints; the prescribed scalable algorithmic structure; and agreement with independent physical benchmarks. For DMRG, the structural requirement is matrix-free application of the effective Hamiltonian. For Pfaffian-MPS, it is tensor-by-tensor construction through Pfaffian overlaps without materializing a dense $2^N$ state in the production algorithm. Where permitted by the run protocol, dense small-system calculations are used only as validation oracles and remain separate from production construction. By contrast, the direct Pfaffian-MPS baseline prohibits dense $2^N$ state vectors at every stage (Sec.~\ref{sec:pfaffian-baseline}). Table~\ref{tab:validation-gates} summarizes the validation gates.

\begin{table*}[tb]
    \centering
    \caption{\textbf{Validation gates for pass/fail classification.} A pass requires every gate specified by the applicable experimental protocol. Run-local specifications, reports, and numerical outputs document the detailed checks, acceptance criteria, and numerical tolerances.}
    \label{tab:validation-gates}
    \renewcommand{\arraystretch}{1.12}
    \setlength{\tabcolsep}{3pt}
    \begin{tabular*}{\textwidth}{@{\extracolsep{\fill}}llp{0.64\textwidth}@{}}
        \toprule
        \textbf{Case} & \textbf{Gate} & \textbf{Criterion} \\
        \midrule
        DMRG & Matrix-free update & The effective Hamiltonian is not materialized; its action is implemented as a linear operator. \\
        DMRG & Heisenberg benchmark & Energy density and entanglement profile are consistent with the Bethe Ansatz and conformal-field-theory validation targets. \\
        DMRG & AKLT benchmark & Energy, bulk bond entropy, and string order are consistent with the analytical AKLT values. \\
        \midrule
        Pfaffian-MPS & Gaussian canonicalization & Unitarity, eigenvalue pairing, and Bloch--Messiah reconstruction satisfy run-specified tolerances. \\
        Pfaffian-MPS & Pfaffian/MPS consistency & Antisymmetry and Pfaffian-identity checks, MPS normalization, and fermion-sign checks pass before production benchmarks. \\
        Pfaffian-MPS & Observable provenance & Reported $E_{\text{MPS}}$ is computed by contracting the constructed MPS; the exact correlation-matrix energy serves as a separate reference. \\
        Pfaffian-MPS & Production benchmark & $N=60$ Kitaev-chain benchmarks satisfy protocol-specific energy-error and convergence requirements (Secs.~\ref{sec:pfaffian-algorithm} and \ref{sec:pfaffian-baseline}), with $\delta_K$ denoting the energy error per site and no dense $2^N$ state materialized in the production algorithm. \\
        \bottomrule
    \end{tabular*}
\end{table*}

The models generated the measurement routines, and the human PI evaluated their outputs against physical benchmarks. Some analytical values, including the AKLT energy $E=-2(L-1)/3$ and string-order plateau $-4/9$, also appeared in the supplied specifications as explicit validation targets. Independence rests on evaluating observables from the generated state and comparing them with analytical results or a separately computed exact reference, as in the Kitaev benchmark. A pass requires both agreement with the physical benchmarks and satisfaction of the architecture and provenance gates.

Outcome denominators follow the experimental units documented in the archive. The DMRG guided grid counts one development path for each specification--implementation model pairing. A path may span multiple conversations, including continuations that receive earlier transcripts and code; unsuccessful intermediate deliveries remain within that path rather than counting as separate grid entries. The 16/16 result therefore denotes successful completion of the 16 tested pairing paths. Direct-baseline and prose-specification totals instead count distinct task attempts. The Pfaffian-MPS roster comprises 26 workflow attempts and includes unsuccessful implementations as well as the Kimi~2.6 self-pairing attempt that ended before code delivery. API failures, context-window failures, and service interruptions before code generation are recorded separately and do not enter the task-outcome denominator.

For descriptive context, we report observed success frequencies with 95\% confidence intervals calculated using the Wilson score method~\cite{Brown2001Binomial}. These intervals quantify finite-sample uncertainty under a binomial model of independent pass/fail trials. Because runs can share specifications, models, prompts, or harnesses, the frequencies and intervals characterize the tested configurations rather than broader model populations.

\subsection{Reusable Protocol and Development Records}

The \emph{Paper-to-Program Many-Body} agent skill packages the workflow as a reusable set of prompts, review procedures, and reporting requirements. It includes the extraction and specification prompts, review checklist, implementation instructions, reporting template, and closed-world constraints~\cite{yizhou_dmrg_llm_2026}. A researcher supplies the source paper and scientific supervision, while one or more agents execute the staged process. The resulting specifications and run records support comparison among implementations, refinement of computational requirements, and evaluation of additional models.

The archive preserves model and harness labels, prompts, transcripts, specifications, code, interventions, and reports. Model labels record the provider snapshots accessed during the study, while run records include dates, harness metadata, and numerical seeds where available. Generation settings are documented when exposed by the interfaces, which did not provide a common temperature, model-side random seed, or token budget. These records support independent examination of the programming process and reproduction of the validated calculations.

\section{Case Study I: DMRG}
\label{sec:dmrg}

The DMRG case study evaluates specification-guided implementation of a two-site, matrix-free algorithm reconstructed from Schollw{\"o}ck's review~\cite{Schollwock2011}. Its detailed equations, pseudocode, and worked examples provide a calibration source for comparing direct implementation with the original three-stage workflow, LLM-0 $\to$ LLM-1 $\to$ LLM-2, under human supervision.

Both conditions target the spin-$1/2$ Heisenberg chain and the spin-$1$ Affleck--Kennedy--Lieb--Tasaki (AKLT) model using NumPy and SciPy. The common implementation objective is matrix-free application of the effective Hamiltonian with Lanczos iteration. Successful implementations must satisfy the physical and computational-architecture gates defined in Sec.~\ref{sec:protocol}. We first compare the direct and specification-guided outcomes, then examine physical and computational validation and the role of specification format.

\subsection{Direct-Implementation Baseline}
\label{sec:dmrg-baseline}

We evaluated four direct-implementation configurations, labeled Gemini 3.1 Pro Preview, GPT~5.4, Claude Opus~4.6, and Kimi Agent in the archived records. Each received Schollw{\"o}ck's review and the same two-site DMRG task, but no intermediate technical specification.

Six of 13 direct attempts pass (46\%; 95\% confidence interval approximately 23--71\%). Gemini succeeds in all three attempts after 1--3 HITL rounds, and Claude succeeds in three of four after 7--11 rounds. GPT and Kimi obtain no passes in three attempts each. Their unsuccessful runs repeatedly materialize the effective Hamiltonian, leading to excessive memory requirements. Successful debugging trajectories generally preserve validated components while correcting new errors, whereas unsuccessful trajectories repeatedly revisit index or contraction inconsistencies. Table~\ref{tab:zeroshot-dmrg} summarizes the outcomes and HITL-round ranges.

\begin{table}[tb]
    \centering
    \caption{\textbf{Direct DMRG baseline.} Implementation proceeds directly from Schollw{\"o}ck's review without an intermediate specification. HITL rounds count feedback after the initial code delivery. The stopping guideline was approximately 20 HITL rounds without convergence, with earlier termination when a model declared that it could not complete the task; the unsuccessful Claude attempt continued to 32 rounds.}
    \label{tab:zeroshot-dmrg}
    \renewcommand{\arraystretch}{1.3}
    \begin{tabular*}{\columnwidth}{@{\extracolsep{\fill}}lccc@{}}
        \toprule
        & & \multicolumn{2}{c}{\textbf{HITL rounds}} \\
        \cmidrule(l){3-4}
        \textbf{Configuration} & \textbf{Pass/Total} & \textbf{Pass} & \textbf{Fail} \\
        \midrule
        Gemini 3.1 Pro & 3/3 & 1--3 & --- \\
        Claude Opus 4.6 & 3/4 & 7--11 & 32 \\
        GPT 5.4 & 0/3 & --- & $>$20 \\
        Kimi Agent & 0/3 & --- & $>$20 \\
        \midrule
        \textbf{Overall} & \textbf{6/13} & & \\
        \bottomrule
    \end{tabular*}
\end{table}

\subsection{Specification-Guided Implementation Across Models}
\label{sec:dmrg-spec}

The specification-guided experiment begins from a common extraction document, \path{main-MPS-Kimi-318.tex}, generated by Kimi~2.5 from the source review. Four specification models refine this shared draft, and each resulting specification is supplied to four implementation configurations, yielding 16 specification--implementation pairings. Fixing the extraction document provides a common starting point for comparing specification development and cross-model transfer.

All 16 specification--implementation development paths satisfy the physics and computational-architecture validation gates (Table~\ref{tab:reproducibility-dmrg}). For the observed 16/16 success frequency, the Wilson 95\% confidence interval has a lower endpoint of approximately 81\% (Sec.~\ref{sec:protocol}). The GPT and Kimi implementation configurations succeed with every tested specification after obtaining no passes in the direct baseline.

\begin{table}[!tbp]
    \centering
    \caption{\textbf{Specification-guided DMRG.} All 16 specification--implementation development paths pass the validation gates, with continuations counted as described in Sec.~\ref{sec:protocol}. Rows identify the specification model (LLM-1), and columns identify the implementation configuration (LLM-2). Parentheses give the recorded HITL rounds after the initial code delivery. Kimi Agent debugs autonomously, with no subsequent human feedback.}
    \label{tab:reproducibility-dmrg}
    \renewcommand{\arraystretch}{1.3}
    \footnotesize
    \setlength{\tabcolsep}{2.5pt}
    \begin{tabular*}{\columnwidth}{@{\extracolsep{\fill}}lcccc@{}}
        \toprule
        & \multicolumn{4}{c}{\textbf{Implementation (LLM-2)}} \\
        \cmidrule{2-5}
        \textbf{Specification (LLM-1)} & Kimi Agent & Gemini & GPT & Claude \\
        \midrule
        \textbf{Kimi}   & $\checkmark\;(0)$ & $\checkmark\;(6)$ & $\checkmark\;(7)$ & $\checkmark\;(1)$ \\
        \textbf{Gemini} & $\checkmark\;(0)$ & $\checkmark\;(5)$ & $\checkmark\;(7)$ & $\checkmark\;(7)$ \\
        \textbf{GPT}    & $\checkmark\;(0)$ & $\checkmark\;(7)$ & $\checkmark\;(13)$ & $\checkmark\;(7)$ \\
        \textbf{Claude} & $\checkmark\;(0)$ & $\checkmark\;(6)$ & $\checkmark\;(10)$ & $\checkmark\;(11)$ \\
        \bottomrule
    \end{tabular*}
\end{table}

Among configurations using HITL debugging, recorded feedback ranges from 1 to 13 rounds: 5--7 for Gemini, 7--13 for GPT, and 1--11 for Claude. Kimi Agent instead debugs autonomously and receives no human feedback after the initial code delivery. All four specifications therefore transfer successfully across the tested implementation configurations, with model-dependent debugging trajectories and feedback requirements.

\subsection{Physical and Computational Validation}
\label{sec:dmrg-validation}

We validate both physical observables and computational architecture. All benchmarks use open chains of $L$ sites. For the spin-$1/2$ Heisenberg chain, the Hamiltonian is $H_{\rm H}=\sum_{i=1}^{L-1}\mathbf{S}_i\cdot\mathbf{S}_{i+1}$, with unit exchange and no applied field. For the spin-$1$ AKLT chain, it is $H_{\rm AKLT}=\sum_{i=1}^{L-1}[\mathbf{S}_i\cdot\mathbf{S}_{i+1}+\frac{1}{3}(\mathbf{S}_i\cdot\mathbf{S}_{i+1})^2]$, with no additive constant. These conventions determine the energy normalization used below.

Physical validation uses ground-state energy, entanglement, and nonlocal order. For the Heisenberg chain, the reference values are the thermodynamic-limit Bethe Ansatz energy density $e_\infty\simeq-0.4431$ and the bipartite-entanglement profile of a Tomonaga--Luttinger liquid with central charge $c=1$. For the AKLT chain~\cite{AKLT1987}, the analytical targets are $E_0=-2(L-1)/3$, bulk bond entropy $\ln 2$, and long-range string order~\cite{denNijs1989,Kennedy1992}. We define
\[
O_{\rm string}^{z}(i,j)=
\left\langle S_i^z \exp\left(\mathrm{i}\pi\sum_{k=i+1}^{j-1}S_k^z\right)S_j^z
\right\rangle,
\]
for which the thermodynamic-limit AKLT plateau is $-4/9$. These observables test both the energy and correlation structure of the computed states.

DMRG assessment combines quantitative energy and tensor-identity checks at run-local tolerances with physical-profile assessment by the human PI. The latter examines finite-size convergence, entanglement-profile shape, and convergence toward the AKLT bulk-entropy and string-order limits. Specifications and transcripts record the numerical settings. These profile-based diagnostics evaluate consistency across system sizes, bonds, or separations rather than imposing a single scalar threshold on every observable.

\begin{figure*}[tb]
    \centering
    \includegraphics[width=0.98\textwidth]{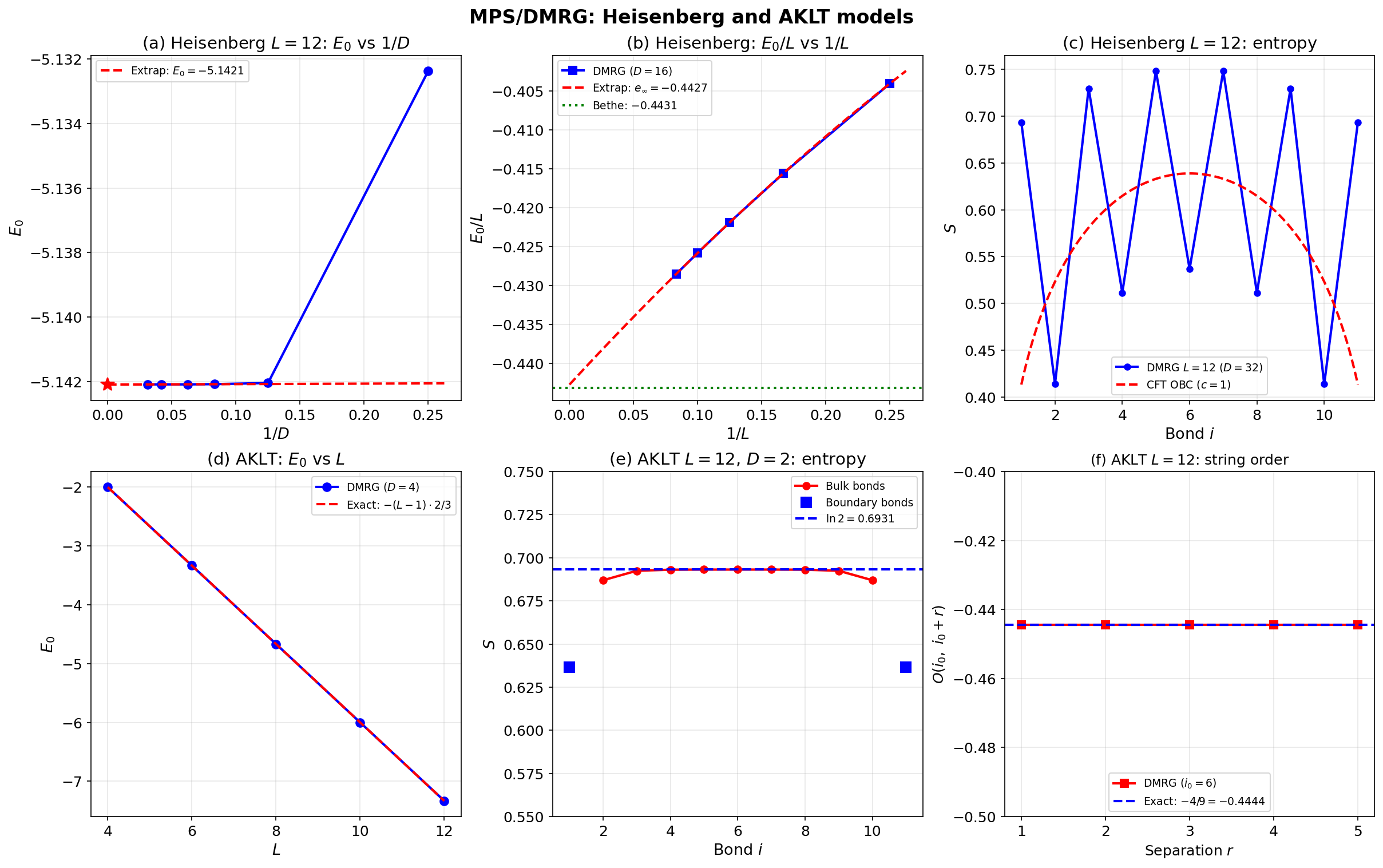}
    \caption{\textbf{DMRG physical validation.} \textbf{(a)}~Ground-state energy versus inverse bond dimension for the $L=12$ Heisenberg chain, with extrapolation to $1/D=0$. \textbf{(b)}~Finite-size extrapolation gives $e_\infty=-0.4427$, compared with the Bethe Ansatz value $-0.4431$. \textbf{(c)}~Entanglement entropy for the open $L=12$ Heisenberg chain, showing boundary-induced even--odd oscillations around the $c=1$ conformal-field-theory reference. \textbf{(d)}~AKLT ground-state energies agree with $E_0=-2(L-1)/3$ across system sizes. \textbf{(e)}~For $L=12$ and bond dimension $D=2$, the bulk bond entropies approach $\ln 2\simeq0.6931$. \textbf{(f)}~AKLT string-order correlations reproduce the $-4/9$ plateau under the convention defined in the text.}
    \label{fig:physics-dmrg}
\end{figure*}

Figure~\ref{fig:physics-dmrg} illustrates these checks for a representative implementation generated by Claude from Gemini's specification; Table~\ref{tab:reproducibility-dmrg} reports the full cross-model outcomes. For the $L=12$ Heisenberg chain, the ground-state energy converges with increasing bond dimension [Fig.~\ref{fig:physics-dmrg}(a)]. A separate finite-size sweep gives $e_\infty=-0.4427$, close to the Bethe Ansatz value $-0.4431$ [Fig.~\ref{fig:physics-dmrg}(b)]. The entanglement profile resolves boundary-induced even--odd oscillations around the $c=1$ conformal-field-theory reference [Fig.~\ref{fig:physics-dmrg}(c)]. The AKLT calculations reproduce the analytical ground-state energy across system sizes, the bulk bond entropy $\ln 2\simeq0.6931$, and the string-order plateau $-4/9$ [Fig.~\ref{fig:physics-dmrg}(d)--(f)].

Computational-architecture validation checks whether the implementation follows the prescribed scalable algorithm. For a representative Heisenberg calculation with $L=20$, $d=2$, and $D_L=D_R=40$, the two-site effective Hamiltonian acts on a space of dimension $N_{\rm loc}=d^2D_LD_R=6400$. Materializing this operator as a dense \texttt{complex128} matrix would require $6400^2\times16\simeq6.6\times10^8$ bytes, or about 0.65~GB (625~MiB). The prescribed matrix-free \texttt{LinearOperator} instead stores contraction environments and applies the Hamiltonian through sequential $\mathcal{O}(D^3)$ contractions. Validation therefore tests computational scalability alongside physical accuracy.

\subsection{Technical Content versus Document Format}
\label{sec:dmrg-ablation}

The successful specification-guided runs motivate a further question: does useful computational guidance depend on mathematical typesetting? GPT~5.4 provides an informative comparison because it succeeds with all four tested specifications after obtaining no direct-baseline passes. In a separate conversion step, GPT~5.4 rewrites one successful specification as informal English prose. The prompt requires preservation of every index convention, tensor shape, contraction sequence, and architectural constraint while removing mathematical typesetting, numbered equations, and structured notation. The resulting prose specification is then supplied to GPT~5.4 in two implementation attempts, each with a budget of 20 HITL rounds.

Both prose-specification attempts pass, requiring 4 and 7 HITL rounds (Table~\ref{tab:ablation-dmrg}). Each implementation satisfies the matrix-free requirement and avoids the dense effective-Hamiltonian construction observed in GPT~5.4's direct-baseline attempts. These results show that preserving the specification's technical content in prose is sufficient to support validated implementation without formal \LaTeX{} presentation.

\begin{table}[tb]
    \centering
    \caption{\textbf{Technical content versus document format.} All conditions use GPT~5.4 for implementation. The direct and formal-specification rows summarize three baseline attempts (Table~\ref{tab:zeroshot-dmrg}) and four implementations from different specifications (Table~\ref{tab:reproducibility-dmrg}), respectively. The prose-specification row reports two attempts using a prose conversion of one successful GPT specification, with all technical content retained. HITL rounds count feedback after the initial code delivery.}
    \label{tab:ablation-dmrg}
    \renewcommand{\arraystretch}{1.3}
    \footnotesize
    \setlength{\tabcolsep}{1.5pt}
    \begin{tabular*}{\columnwidth}{@{\extracolsep{\fill}}lccc@{}}
        \toprule
        \textbf{Condition} & \textbf{Input} & \shortstack{\textbf{Pass/}\\\textbf{Total}} & \shortstack{\textbf{HITL}\\\textbf{rounds}} \\
        \midrule
        Direct & Review article & 0/3 & $>$20 \\
        Formal specification & \LaTeX{} document & 4/4 & 7--13 \\
        Prose specification & English prose & 2/2 & 4, 7 \\
        \bottomrule
    \end{tabular*}
\end{table}

The DMRG case establishes successful cross-model specification transfer and shows that its technical guidance remains effective in prose form. We next examine specification transfer and refinement for a compact-source algorithm involving fermionic canonicalization and Pfaffian overlaps.

\section{Case Study II: Pfaffian-MPS}
\label{sec:pfaffian}

The second case study examines the Pfaffian-based conversion of Hartree--Fock--Bogoliubov (HFB) wave functions into MPS described in the source Letter~\cite{Jin2022}. This five-page article presents the method without accompanying source code, providing a compact and technically demanding implementation task. Unlike the variational sweeps of DMRG, the method constructs MPS tensors from fermionic Gaussian-state overlaps and requires consistent treatment of canonicalization, operator ordering, phases, and fermionic signs. We apply the expanded workflow described in Sec.~\ref{sec:methodology}, including dedicated specification review and structured implementation reporting, with NumPy, SciPy, and Matplotlib as the permitted dependencies.

The author coauthored the source Letter and draws on this domain expertise to define validation gates and evaluate physical diagnostics. Human supervision during implementation follows the protocol in Sec.~\ref{sec:llm2}, focusing on physical observables and runtime behavior while leaving computational diagnosis and code revision to the implementation agent. The archived transcripts preserve these interventions together with the generated code and validation results.

\subsection{Algorithm and Benchmark Design}
\label{sec:pfaffian-algorithm}

We implement the Letter's Pfaffian construction rather than the matrix-product-operator-based conversion included there for comparison. Starting from the $2N\times2N$ correlation matrix of a fermionic Gaussian state, the algorithm constructs its MPS representation tensor by tensor. At each chain cut, the Schmidt vectors of the left subsystem are expressed as Bogoliubov vacua in a suitable canonical basis, and the MPS tensor elements are obtained from overlaps evaluated with a Pfaffian formula. A consistent implementation combines Bloch--Messiah decomposition, particle--hole transformations for occupied Schmidt modes, stable canonicalization near entanglement eigenvalues $1/2$, Pfaffian block ordering, and Jordan--Wigner signs.

Consistent phase conventions and operator ordering are required to connect the canonical decompositions at successive chain cuts. At production scale, reusing intermediate contractions avoids repeated work in evaluating the overlaps. The production benchmarks therefore test both physical accuracy and tensor-by-tensor MPS construction at $N=60$.

We benchmark the method using the Kitaev-chain Hamiltonian in Eq.~(13) of Ref.~\cite{Jin2022}, with hopping $t=1$, chemical potential $\mu=0.5$, and antiperiodic boundary conditions. We define the absolute energy error per site as
\begin{equation}
    \delta_K=\frac{|E_{\mathrm{MPS}}-E_{\mathrm{exact}}|}{N},
\end{equation}
where $E_{\mathrm{MPS}}$ is obtained by contracting the constructed MPS and $E_{\mathrm{exact}}$ is the independently computed exact ground-state energy for the same Hamiltonian and boundary conditions. This definition directly tests the physical accuracy of the generated MPS.

The specification-guided benchmark schedule covers eight pairing strengths, $\Delta=0,\ 0.025,\ 0.05,\ 0.1,\ 0.2,\ 0.4,\ 0.6,\ 0.8$, with $\chi=256$ for $\Delta<0.2$ and $\chi=64$ for $\Delta\geq0.2$. A workflow pass requires $\delta_K<10^{-5}$ at every prescribed point with $\Delta\geq0.2$, together with the mathematical and provenance gates in Table~\ref{tab:validation-gates}. The lower-pairing points probe convergence and computational performance. Run reports record completed points and any changes to bond dimensions or truncation cutoffs; when runtime limits the full low-pairing schedule, they identify the reduced sweep explicitly.

\subsection{Direct-Implementation Baseline}
\label{sec:pfaffian-baseline}

In the direct baseline, the implementation agent receives the source Letter without an intermediate technical specification. The prompt requires Bloch--Messiah decomposition, Pfaffian/Wick-based overlap evaluation, per-tensor cost polynomial in $(N,\chi)$, and energy evaluation by contracting the constructed MPS. It designates the Letter as the sole physics reference and permits only NumPy, SciPy, and Matplotlib as dependencies. The prompt also prohibits dense $2^N$ state vectors at every stage and requires a small-system brute-force overlap check before the production benchmark. Implementation and debugging proceed autonomously for up to 30 auto-debug rounds.

The direct benchmark evaluates an $N=60$ chain with $\chi=64$ at six pairing strengths, $\Delta=0.025,\ 0.05,\ 0.1,\ 0.2,\ 0.5,\ 0.8$. A pass requires $\delta_K<10^{-5}$ at every prescribed point with $\Delta\geq0.2$ and $\delta_K<10^{-2}$ at $\Delta=0.025$. At $\Delta=0.025$, it further requires convergence of the form $\delta_K\propto(1/\chi)^\alpha$ with $\alpha>1$ over $\chi\in\{16,32,64\}$. Thus, low-pairing accuracy and convergence enter the formal criteria for the direct baseline, whereas in the specification-guided workflow they serve as additional performance diagnostics.

The direct baseline comprises eleven attempts using the same model--harness combinations as the Phase~1 self-pairing experiments listed in Table~\ref{tab:pfaffian-runs}. None satisfies the full audit and validation criteria (0/11). The archived records identify obstacles including incomplete $N=60$ calculations, use of correlation-matrix energies in place of measurements from the constructed MPS, and unresolved normalization or fermion-sign checks. The specification-guided experiments test whether reviewed technical guidance can address these obstacles and support validated tensor construction across implementation agents.

\subsection{Specification-Guided Experiments}
\label{sec:pfaffian-workflow}

The specification-guided experiments address three questions. First, can a model implement the specification it develops? Second, how effectively do specifications transfer across models, and what do the outcomes reveal about the implementation agents? Third, can lessons from earlier attempts be consolidated into guidance that supports previously unsuccessful configurations? These questions define three phases: model self-pairing, cross-model specification transfer, and implementation from a consolidated specification. Tables~\ref{tab:pfaffian-runs}, \ref{tab:phase2-pfaffian}, and~\ref{tab:phase3-pfaffian} identify the specification source, implementation model, and harness for all 26 workflow attempts.

\subsubsection{Phase 1: Model Self-Pairing}
\label{sec:pfaffian-phase1}

In Phase~1, the same foundation model generates the specification and implements the code in eleven workflow attempts. Four attempts pass: GPT~5.5 under Codex and OpenCode, and Claude Opus~4.8 under ClaudeCode and OpenCode (Table~\ref{tab:pfaffian-runs}). DeepSeek V4 Pro and Gemini 3.1 Pro do not pass under either Codex or OpenCode. The Kimi~2.6 attempt under KimiCode ends before code delivery and remains in the workflow denominator. Within the partially sampled model--harness grid, each model obtains the same outcome under every tested harness.

\begin{table}[!tbp]
    \centering
    \caption{\textbf{Model self-pairing (Phase~1).} The same foundation model generates the specification and implements the code. Pass classifications follow the workflow criteria in Sec.~\ref{sec:pfaffian-algorithm}. ``No code delivered'' counts as an unsuccessful workflow attempt.}
    \label{tab:pfaffian-runs}
    \renewcommand{\arraystretch}{1.12}
    \footnotesize
    \begin{tabular*}{\columnwidth}{@{\extracolsep{\fill}}llc@{}}
        \toprule
        \textbf{Model (LLM-1 = LLM-2)} & \textbf{Harness} & \textbf{Status} \\
        \midrule
        GPT 5.5 & Codex & Pass \\
        GPT 5.5 & OpenCode & Pass \\
        Claude Opus 4.8 & ClaudeCode & Pass \\
        Claude Opus 4.8 & OpenCode & Pass \\
        Claude Opus 4.7 & ClaudeCode & Fail \\
        DeepSeek V4 Pro & OpenCode & Fail \\
        DeepSeek V4 Pro & Codex & Fail \\
        Gemini 3.1 Pro & OpenCode & Fail \\
        Gemini 3.1 Pro & Codex & Fail \\
        Kimi 2.6 & KimiCode & No code delivered \\
        GPT 5.4 & Codex & Fail \\
        \textbf{Phase 1 overall} & & \textbf{4/11} \\
        \bottomrule
    \end{tabular*}
\end{table}

The unsuccessful attempts identify two recurring challenges: preserving fermion-sign and gauge consistency, and completing the $N=60$ construction within the available memory and runtime. Because self-pairing assigns specification development and implementation to the same model, Phase~2 tests whether transferring these specifications to different implementation agents supports validated construction.

\subsubsection{Phase 2: Cross-Specification Transfer}
\label{sec:pfaffian-phase2}

Phase~2 evaluates cross-model specification reuse in eight workflow attempts. GPT~5.5 under Codex passes with all four specifications generated by non-GPT models. In the reverse direction, none of the four tested non-GPT implementation models passes with a GPT~5.5 specification (Table~\ref{tab:phase2-pfaffian}).

\begin{table*}[tb]
    \centering
    \caption{\textbf{Cross-specification transfer (Phase~2).} GPT~5.5 under Codex achieves 4/4 passes with non-GPT specifications, compared with 0/4 passes when the four tested non-GPT models implement from GPT~5.5 specifications. Each row identifies the specification source, implementation model, and harness. Pass classifications follow the workflow criteria in Sec.~\ref{sec:pfaffian-algorithm}.}
    \label{tab:phase2-pfaffian}
    \renewcommand{\arraystretch}{1.15}
    \begin{tabular*}{\textwidth}{@{\extracolsep{\fill}}llllc@{}}
        \toprule
        \textbf{Direction} & \textbf{Spec source} & \textbf{Implementation model} & \textbf{Harness} & \textbf{Status} \\
        \midrule
        Non-GPT spec $\to$ GPT & Claude Opus 4.7 & GPT 5.5 & Codex & Pass \\
        Non-GPT spec $\to$ GPT & DeepSeek V4 Pro & GPT 5.5 & Codex & Pass \\
        Non-GPT spec $\to$ GPT & Gemini 3.1 Pro & GPT 5.5 & Codex & Pass \\
        Non-GPT spec $\to$ GPT & Kimi 2.6 & GPT 5.5 & Codex & Pass \\
        \midrule
        GPT spec $\to$ non-GPT & GPT 5.5 & Claude Opus 4.7 & ClaudeCode & Fail \\
        GPT spec $\to$ non-GPT & GPT 5.5 & DeepSeek V4 Pro & Codex & Fail \\
        GPT spec $\to$ non-GPT & GPT 5.5 & Gemini 3.1 Pro & Codex & Fail \\
        GPT spec $\to$ non-GPT & GPT 5.5 & Kimi 2.6 & Codex & Fail \\
        \bottomrule
    \end{tabular*}
\end{table*}

The successful transfers show that specifications from models with unsuccessful self-pairing attempts support validated construction by another implementation agent. The reverse-direction attempts encountered fermion-sign errors, incomplete $N=60$ calculations, unresolved Bloch--Messiah reconstruction, and use of exact-energy proxies instead of MPS measurements. This directional asymmetry makes specification reuse diagnostically useful: implementation outcomes depend jointly on the supplied technical guidance and the implementation agent's ability to realize it.

\subsubsection{Phase 3: Consolidated Specification}
\label{sec:pfaffian-phase3}

Phase~3 tests whether implementation lessons can be incorporated into a consolidated specification that supports subsequent validated construction. Building on the earlier agent-reviewed specification, GPT~5.5 under Codex synthesizes specifications generated by GPT~5.5, GPT~5.4, Claude Opus~4.7, DeepSeek V4 Pro, and Kimi~2.6 together with implementation reports and debugging evidence, producing \path{latex-spec-LLM-1-Final.tex}. The consolidation proceeds without an additional human review pass. The resulting document expands the guidance on fermionic signs, zero-mode handling, pair-phase conventions, and reuse of intermediate contractions. Implementation agents receive both \path{latex-spec-LLM-1.tex} and the consolidated document, with the latter taking precedence wherever they differ.

Implementation feedback continued to refine the consolidated document during Phase~3. The preserved KimiCode/Kimi~2.6 copy does not include the later normalization-debug addendum. The Codex/DeepSeek V4 Pro copy includes that addendum but not the subsequent clarifications of Bloch--Messiah record identity and occupied-sector gauge fixing. The other five run-local copies contain all three additions. These files record the revision state preserved with each implementation; because the same filename was retained across revisions, they do not by themselves establish the exact initial input for every run. Phase~3 therefore evaluates both reuse and iterative refinement of the consolidated guidance, with the development records providing its revision context.

With the consolidated specification, three of seven workflow attempts pass: DeepSeek V4 Pro and GPT~5.4 under Codex, and Claude Opus~4.7 under ClaudeCode (Table~\ref{tab:phase3-pfaffian}). Each of these model--harness configurations was unsuccessful in Phase~1 self-pairing. Gemini 3.1 Pro and Kimi~2.6 remain unsuccessful in the tested configurations, with recurring canonicalization or proxy-observable issues. Together, Phases~2 and~3 show that expanded computational guidance enables previously unsuccessful configurations to pass, while implementation capability continues to shape the outcomes.

\begin{table}[!tbp]
    \centering
    \caption{\textbf{Consolidated specification (Phase~3).} All runs use \texttt{latex-spec-\allowbreak LLM-1-Final.tex} as the controlling reference, with run-local revisions described in the text. Pass classifications follow the workflow criteria in Sec.~\ref{sec:pfaffian-algorithm}. The three passing model--harness configurations were unsuccessful in Phase~1.}
    \label{tab:phase3-pfaffian}
    \renewcommand{\arraystretch}{1.12}
    \footnotesize
    \begin{tabular*}{\columnwidth}{@{\extracolsep{\fill}}llc@{}}
        \toprule
        \textbf{Implementation (LLM-2)} & \textbf{Harness} & \textbf{Status} \\
        \midrule
        DeepSeek V4 Pro & Codex & Pass \\
        GPT 5.4 & Codex & Pass \\
        Claude Opus 4.7 & ClaudeCode & Pass \\
        Gemini 3.1 Pro & Codex & Fail \\
        Kimi 2.6 & Codex & Fail \\
        Kimi 2.6 & KimiCode & Fail \\
        Gemini 3.1 Pro & OpenCode & Fail \\
        \textbf{Phase 3 overall} & & \textbf{3/7} \\
        \bottomrule
    \end{tabular*}
\end{table}

Across the three phases, 11 of 26 workflow attempts pass: 4/11 in model self-pairing, 4/8 in cross-specification transfer, and 3/7 with the consolidated specification. This corresponds to an overall success frequency of approximately 42\%, with a 95\% Wilson confidence interval of approximately 26--61\% (Sec.~\ref{sec:protocol}). The staged experiments produce multiple validated implementations and document how cross-model transfer and iterative specification refinement contribute to their development.

\subsection{Physics Validation}
\label{sec:pfaffian-physics}

\begin{figure*}[tb]
    \centering
    \includegraphics[width=0.72\textwidth]{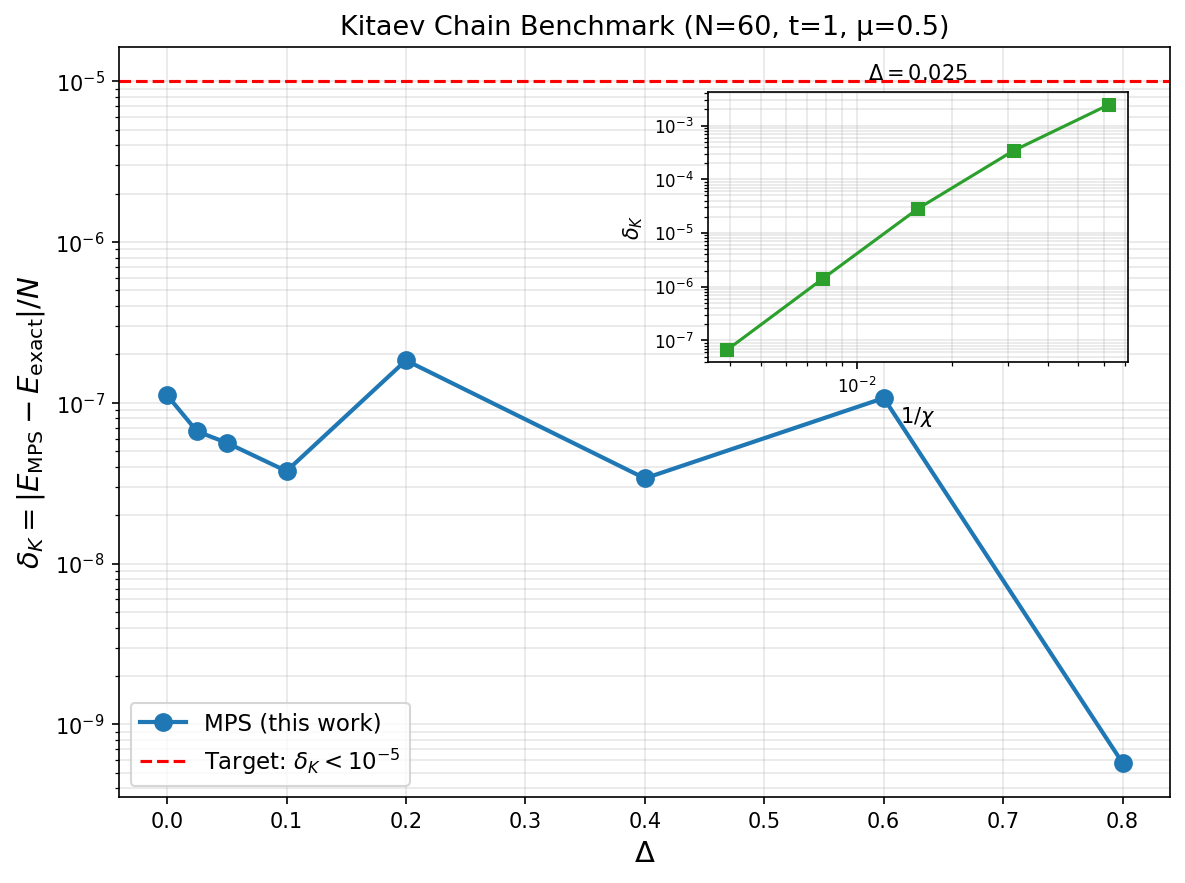}
    \caption{\textbf{Pfaffian-MPS physics validation.} Representative Phase~3 results from Claude Opus~4.7 under ClaudeCode using the consolidated specification, for $N=60$, $t=1$, $\mu=0.5$, and antiperiodic boundary conditions. The main panel shows the energy error per site $\delta_K$, with $\chi=256$ for $\Delta<0.2$ and $\chi=64$ for $\Delta\geq0.2$. All eight pairing values lie below the dashed $10^{-5}$ line, the required accuracy threshold for $\Delta\geq0.2$. The inset shows convergence at $\Delta=0.025$ for $\chi\in\{16,32,64,128,256\}$, with approximately power-law error reduction consistent with the source benchmark~\cite{Jin2022}.}
    \label{fig:kitaev}
\end{figure*}

Across the passing attempts, $E_{\mathrm{MPS}}$ is obtained by contracting the constructed MPS, and every prescribed benchmark with $\Delta\geq0.2$ satisfies the required accuracy threshold. Figure~\ref{fig:kitaev} shows the full pairing-strength sweep for a representative Phase~3 implementation generated by Claude Opus~4.7 under ClaudeCode from the consolidated specification. This implementation achieves $\delta_K<10^{-5}$ at all eight pairing values, including the lower-pairing points used as performance diagnostics. At $\Delta=0.025$, the bond-dimension sweep $\chi\in\{16,32,64,128,256\}$ exhibits approximately power-law error reduction, consistent with the source benchmark~\cite{Jin2022}. Approaching the gapless limit increases the entanglement and therefore the bond dimension required for accurate representation.

\subsection{Implementation Lessons and Reusable Validation Rules}
\label{sec:failure-modes}

The archived attempts yield four implementation lessons that translate the observed outcomes into reusable rules for specification design and validation.

\textit{Observable provenance.} Some attempts obtained accurate numerical energies by using the exact correlation-matrix result as $E_{\mathrm{MPS}}$. Computing $E_{\mathrm{MPS}}$ through explicit contraction of the constructed MPS and evaluating the exact reference independently ensures that the comparison tests the generated state. Numerical accuracy and measurement provenance are therefore complementary validation requirements.

\textit{Scale validation.} Passing exact Fock-space tests at $N\leq4$ does not guarantee production-scale behavior: some implementations failed at $N\geq5$ or did not complete the $N=60$ benchmark. Small-system oracles isolate algebraic correctness, while production benchmarks evaluate truncation accuracy, computational organization, and scaling. Requiring both connects local mathematical consistency to practical execution at the intended system size.

\textit{Global conventions.} The recurring fermion-sign and reconstruction errors show that operator ordering, bra/ket conventions, particle--hole transformations, and Bloch--Messiah pair phases must be fixed consistently throughout the chain. Recording these choices at the algorithm level and validating the corresponding identities before assembling larger MPS tensors make consistency across subsystem cuts explicit and testable.

\textit{Computational reuse.} Evaluating every tensor element independently may reproduce small-system results but becomes impractical at $N=60$ and $\chi=256$. Reusing intermediate contractions, with explicit accounting of memory and runtime costs, turns the Pfaffian overlap formula into a production-scale algorithm. Like matrix-free updates in DMRG, this computational organization is an integral implementation requirement rather than an optional optimization.

These lessons convert implementation experience into explicit specification clauses and validation gates that connect algebraic correctness, observable provenance, and production-scale performance.

\section{Discussion}
\label{sec:discussion}
\label{sec:synthesis}

\subsection{Externalized Knowledge and Implementation Capability}
\label{sec:two-bottleneck}

The two case studies show that specification-guided programming combines two components: externalized computational knowledge and the implementation agent's ability to realize it. A reviewed specification serves as a reusable intermediate representation that makes equations, conventions, and computational requirements explicit, guides implementation and debugging, and provides a common reference for interpreting differences in outcomes across agents.

The DMRG results demonstrate the constructive role of explicit specifications. All 16 tested specification--implementation model pairings pass, including implementation configurations that obtain no pass in the direct baseline. Both prose-specification attempts also pass, showing that essential computational guidance is retained without mathematical typesetting. Explicit tensor shapes, contraction sequences, and matrix-free architecture provide agents with an actionable framework for implementation and debugging.

The Pfaffian-MPS results reveal the diagnostic role of specification reuse. Cross-model transfer is asymmetric: GPT~5.5 under Codex passes with all four tested non-GPT specifications, whereas none of the four tested non-GPT implementation models passes with a GPT~5.5 specification. The consolidated specification subsequently supports three model--harness configurations that were unsuccessful in Phase~1. These results support a \emph{two-bottleneck picture}: one bottleneck lies in externalizing the source method as complete computational guidance, and the other in the implementation agent and execution environment realizing and debugging that guidance. Phase~3 shows how these bottlenecks interact, as richer guidance broadens the set of successful implementations without removing dependence on implementation capability.

The implementation records further illustrate this interaction. Several agents identify and correct inconsistencies in the supplied specifications while satisfying structural validation checks. This feedback loop makes a reviewed specification a revisable computational reference: validation tests the program and identifies guidance that requires refinement. Successful Pfaffian-MPS implementations also differ in module organization, overlap strategy, and treatment of fermionic signs, showing that the specification defines required invariants without prescribing a single code architecture.

This picture identifies two complementary paths for improvement. Ambiguities shared across agents identify conventions and computational details that should be made explicit. Divergent outcomes from reused specifications motivate closer examination of implementation-level reasoning, debugging strategies, and computational organization. Incorporating both kinds of evidence into revised specifications converts development experience into reusable computational knowledge.

\subsection{Provenance and Reproducible Scientific Outputs}
\label{sec:contamination}

The archived prompts, transcripts, specifications, and code make the development process traceable by documenting its scientific inputs, computational choices, and human interventions. Under the closed-world protocol, agents receive the supplied literature or specifications and use the permitted dependencies without retrieving or being supplied with existing tensor-network implementations. These records establish session-level provenance. They do not determine whether the models encountered related software or mathematical routines during pretraining, because the training corpora are unavailable for inspection.

The archived implementations and validation records support reproduction of the reported calculations: readers can rerun the programs, verify the physical observables, and trace each measurement to its computational procedure. These standalone implementations complement established software such as ITensor, TeNPy, and TeMFpy by exposing a dependency-constrained path from the source method to validated numerical results.

Repeating the programming experiment is distinct from rerunning a completed calculation. The same prompts initiate a new stochastic development trajectory whose outcome can vary with the model and harness. Versioned specifications, time-stamped development records, and explicit validation criteria provide fixed reference points for comparing new attempts. The archive therefore supports both reproduction of completed calculations and systematic evaluation of the process that generates them.

\subsection{Scientific Supervision and Development Effort}
\label{sec:pedagogical}

The workflow assigns complementary responsibilities to the agents and the researcher. Agents develop specifications and programs, diagnose implementation errors, and execute tests; the researcher selects the scientific task, defines acceptance criteria, and evaluates physical observables and runtime behavior. During HITL collaboration, feedback communicates scientific and computational objectives while the agent remains responsible for diagnosis and code revision, as described in Sec.~\ref{sec:llm2}. Specifications, transcripts, and validation records make this division of labor traceable.

The recorded development times illustrate how human and agent activity combine. The initial DMRG path from source literature to validated code took less than 24 hours of wall-clock time, including approximately 14 hours of active HITL collaboration; subsequent DMRG paths took less time. Successful Pfaffian-MPS paths required somewhat more elapsed time but comparable active human effort, while additional specification and review cycles ran autonomously. Active collaboration time therefore measures direct human effort, whereas wall-clock time also reflects model latency, queueing, autonomous execution, and human availability.

Scientific supervision also creates value beyond an individual implementation. Physical diagnostics identify issues requiring further investigation, while implementation records translate the resulting lessons into specification clauses and review criteria. Both successful and unsuccessful attempts therefore contribute evidence that improves the guidance available for subsequent programming tasks.

\subsection{Scope and Next Experiments}
\label{sec:limitations}

The two case studies span complementary settings in MPS-based many-body computation: DMRG combines a detailed pedagogical source with variational optimization, whereas Pfaffian-MPS combines a compact research Letter with direct tensor construction from fermionic Gaussian-state overlaps. Because source presentation, algorithmic structure, and workflow design vary together, the results characterize complete task--workflow configurations. They provide a basis for further experiments on implementation configurations, workflow components, transfer across researchers, and broader algorithmic scope.

A factorial study crossing implementation models and harnesses, with repeated trials under matched inference and debugging budgets, would extend the present samples, in which some attempts share specifications, to estimate model, harness, and interaction effects and test the robustness of the transfer asymmetry. Recording active human effort, agent computation, and elapsed time separately would quantify the resources required for validated implementation. A matched manual-implementation comparison would then enable a direct assessment of development speedup.

Component-level experiments could independently vary the computational guidance supplied to the agent and the requirement to execute intermediate validation checks with stop gates. All conditions would receive the same scientific task and undergo a common independent final evaluation. This design would separate the contribution of explicit implementation guidance from that of validation-gated development; these components were refined together in the Pfaffian-MPS workflow. The two successful prose-specification attempts also motivate repeated formal-versus-prose comparisons with independent verification that the technical content is matched.

Transfer across researchers provides another important test. In the present study, the author supplied all human supervision, drawing on prior expertise in both tasks and coauthorship of the source Pfaffian-MPS paper. The restricted-feedback protocol and intervention records make this contribution traceable. Repeating the workflow with researchers outside the source-method team would test the portability of the specifications and supervision protocol across users.

Broader algorithmic tests can assess the generality of the specification-first approach beyond the present case studies. Natural candidates include time-dependent variational principle methods~\cite{Haegeman2011}, infinite MPS~\cite{Vidal2007}, infinite-system DMRG~\cite{McCulloch2008}, projected entangled-pair states~\cite{Verstraete2004}, and hybrid Gutzwiller-guided DMRG~\cite{Jin2021,Jin2025}. This set of methods extends the evaluation to time evolution, infinite systems, higher-dimensional tensor networks, and hybrid computational strategies.

The public workflow skill and versioned archive provide a common foundation for this research agenda. Their prompts, specifications, implementation records, and validation evidence allow future studies to reuse established baselines, introduce controlled variations, and compare outcomes. The present case studies thus provide starting points for cumulative evaluation and refinement of scientific programming workflows.

\section{Conclusion}
\label{sec:conclusion}

Reviewed technical specifications provide a practical intermediate representation between source literature and validated quantum many-body research software. The DMRG study establishes broad cross-model transfer: all 16 specification--implementation development paths pass, compared with 6 of 13 direct attempts, and both prose-specification attempts also pass. The Pfaffian-MPS study shows a more selective pattern: under matched model--harness configurations, 4 of 11 self-paired workflow attempts pass compared with none of 11 direct attempts; across all phases, 11 of 26 workflow attempts pass, cross-model transfer is asymmetric, and the consolidated specification supports three model--harness configurations that were unsuccessful with earlier guidance.

The combined evidence shows that technical specifications play constructive and diagnostic roles: they guide implementation and provide a common reference for tracing differences in outcome to the source presentation or to the implementation agent. Validation creates a feedback loop in which implementation outcomes refine both the computational guidance and the programs. Physical benchmarks, provenance checks, and production-scale tests anchor this process in verifiable scientific outputs. The \emph{Paper-to-Program Many-Body} protocol and public archive package these practices into a reusable framework for extending specification-guided programming to other many-body algorithms.

\begin{acknowledgments}
The author thanks Chen Fang, Kun Jiang, Yuan Li, Zi-Xiang Li, Yuan Wan, and Lei Wang for helpful discussions, and thanks the developers of ITensor and TeNPy, as well as the broader open-source tensor-network community, for their contributions to computational many-body research. This work is supported by the National Key Research and Development Program of China (No.~2022YFA1403403) and the National Natural Science Foundation of China (Nos.~12274441 and 12534004).

Foundation models and agentic harnesses served as experimental systems. The evaluated models were Kimi 2.5 and 2.6 (Moonshot AI), Gemini 3.1 Pro Preview (Google), GPT 5.4 and 5.5 (OpenAI), Claude Opus 4.6, 4.7, and 4.8 (Anthropic), and DeepSeek V4 Pro (DeepSeek). The harnesses were Codex, ClaudeCode, OpenCode, and KimiCode. The author designed the study, selected the source problems, defined the physical-validation and provenance criteria, supervised the experimental runs, audited the implementations, and determined the reported pass/fail classifications.

Generative-AI systems also assisted manuscript preparation through substantive drafting, restructuring, critical revision, and language editing, beyond proofreading or language polishing. The sole human author selected and verified the scientific content, checked the results and claims against the archived evidence, reviewed and revised the manuscript text, and takes full responsibility for the work. No AI system is listed as an author. The writing process for an earlier version is described in the companion essay \textit{Co-Authoring with AI: How I Wrote a Physics Paper About AI, Using AI}~[\href{https://arxiv.org/abs/2604.04081}{arXiv:2604.04081}]. The workflow illustration was generated with Nano Banana 2 from detailed prompts under human supervision; the numerical physics figures were generated from the archived calculations.
\end{acknowledgments}

\section*{Data and Code Availability}

A versioned snapshot of the study materials is archived on Zenodo as \texttt{DMRG-LLM} v3.0.0 under CC BY 4.0~\cite{yizhou_dmrg_llm_2026}, with a corresponding \href{https://github.com/yizhou76-sudo/DMRG-LLM/releases/tag/v3.0.0}{GitHub source release}. The \texttt{DMRG/} and \texttt{Pfaffian-MPS/} collections contain prompts, transcripts, specifications, implementation runs, validation artifacts, figures, and summary records. The Pfaffian-MPS collection additionally includes the reusable workflow skill and consolidated specifications. Each collection provides a README describing its organization and entry points. Source articles are cited in the bibliography; duplicate run-local copies of the Pfaffian source PDF are omitted from the public archive.

\bibliography{references}

\end{document}